\documentstyle[12pt]{article}
\def\beq{\begin{equation}}
\def\eeq{\end{equation}}
\def\bea{\begin{eqnarray}}
\def\eea{\end{eqnarray}}
\def\pr{\prime}
\def\pa{\partial}
\newcommand\be{\begin{equation}}
\newcommand\ee{\end{equation}}
\newcommand{\cl}{\centerline}
\renewcommand{\theequation}{\thesection.\arbic{equation}}
\renewcommand{\theequation}{A.\arbic{equation}}
\begin{document}
\begin{titlepage}
\setlength{\textwidth}{5.0in} \setlength{\textheight}{7.5in}
\setlength{\parskip}{0.0in} \setlength{\baselineskip}{18.2pt}
\setcounter{page}{1}
\begin{flushright}
HD-THEP-03-40\\
\end{flushright}
\vskip1.5cm \cl{\large{\bf The Gauged $O(3)$ Sigma Model:
Schr\"odinger Representation}}\par \cl{\large{\bf and
Hamilton-Jacobi Formulation}}\par \vskip 1.0cm \cl{Soon-Tae
Hong$^{1,2}$ and Klaus D. Rothe$^{2}$} \vskip 0.4cm
\cl{$^{1}$Department of Science Education} \cl{Ewha Womans
University, Seoul 120-750, Korea} \cl{$^{2}$Institut f\"ur
Theoretische Physik, Universit\"at Heidelberg} \cl{Philosophenweg
16, D-69120 Heidelberg, Germany} \vfill
\begin{center}
{\bf ABSTRACT}
\end{center}
\begin{quote}
We first study a free particle on an $(n-1)$-sphere in an extended
phase space, where the originally second-class Hamiltonian and
constraints are now in strong involution. This allows for a
Schr\"odinger representation and a Hamilton-Jacobi formulation of
the model. We thereby obtain the free particle energy spectrum
corresponding to that of a rigid rotator. We extend these
considerations to a modified version of the field theoretical O(3)
nonlinear sigma model, and obtain the corresponding energy
spectrum as well as BRST Lagrangian.
\end{quote}
\end{titlepage}

%%%%%%%%%%%%%%%%%%%%%%%%%%%%%%%%%%%%%%%%%%%%%%%%%%%%%%%%%%%%%%%%%%%%%%%%%%
\section{Introduction}
%%%%%%%%%%%%%%%%%%%%%%%%%%%%%%%%%%%%%%%%%%%%%%%%%%%%%%%%%%%%%%%%%%%%%%%%%%
The quantization of constraint systems has been extensively
discussed in the literature~\cite{dirac64,bft,bft1,hong02pr}. In
particular, the embedding of a second-class system into a
first-class one~\cite{bft}, where constraints are in strong
involution, has been of much interest, and has found a large
number of applications~\cite{bft1,hong02pr}. However, the
usefulness of having the constraints in strong involution has, in
our view, not been sufficiently emphasized. To illustrate this is
the main objective of the present paper. Recently we have obtained
in compact form the first-class Hamiltonian for the
O(3) nonlinear sigma model (NLSM) ~\cite{hong99o3}. On the one
hand, this model is of interest, since it leads to novel
phenomenological aspects~\cite{cs,eucl}. On the other hand, it has
also served to investigate the Lagrangian, symplectic, and
Batalin-Fradkin-Tyutin (BFT) embedding procedure to its
quantization~\cite{rothe03}. In this paper we wish to illustrate
in terms of this model the central role which the BFT embedding
plays for obtaining a Schr\"odinger representation and
Hamilton-Jacobi formulation. In order to clarify the relation
between the first-class formulation and the Schr\"odinger
representation of a constrained quantum mechanical system, we
shall first consider the quantum mechanical analog of a modified
NLSM, i.e. the free particle motion on an $(n-1)$-sphere in an
extended phase space. We then apply the Schr\"odinger
representation approach, developed for the $(n-1)$-sphere, to a
slightly modified version of the gauged, field theoretical O(3)
nonlinear sigma model, within the framwork of the BFT formalism.

%%%%%%%%%%%%%%%%%%%%%%%%%%%%%%%%%%%%%%%%%%%%%%%%%%%%%%%%%%%%%%%%%%%%%%%%%%
\section{Schr\"odinger representation for a free particle on an $(n-1)$-sphere}
\setcounter{equation}{0}
\renewcommand{\theequation}{\arabic{section}.\arabic{equation}}
%%%%%%%%%%%%%%%%%%%%%%%%%%%%%%%%%%%%%%%%%%%%%%%%%%%%%%%%%%%%%%%%%%%%%%%%%%
Consider the motion of a particle on a hypersphere $S^{n-1}$, as
described by the Lagrangian,
\begin{equation}
L_{0}=\frac{1}{2}\dot{q}_{a}\dot{q}_{a}+\lambda q_{a}\dot{q}_{a},
\label{Lag}
\end{equation}
where $q_{a}$ $(a=1,2,...,n)$ are the coordinates parameterizing
the $S^{n-1}$ manifold, and $\lambda$ is the Lagrange multiplier
implementing the second-class constraint $q_{a}\dot{q}_{a}\approx
0$ associated with the geometrical constraint $q_{a}q_{a}={\rm
constant}$.\footnote{It turns out to be convenient not to impose
the constraint in the form $\lambda (q_{a}q_{a}-1)$.} From the
Lagrangian (\ref{Lag}) we obtain for the canonical momenta
conjugate to the multiplier $\lambda$ and the coordinates $q_{a}$
\beq p_{\lambda}=0,~~~ p_{a}=\dot{q}_{a}+\lambda q_{a},
\label{momenta} \eeq and the corresponding canonical Hamiltonian
reads
\begin{equation}
H_{0}=\frac{1}{2}(p_{a}-\lambda q_{a})(p_{a}-\lambda q_{a}).
\label{canH}
\end{equation}
The usual Dirac algorithm is readily shown to lead to the pair of
second-class constraints $\Omega_{i}$ $(i=1,2)$:
\beq\label{Omega12} \Omega_{1}=p_{\lambda}\approx
0,~~~\Omega_{2}=q_{a}p_{a}-\lambda q_{a}q_{a}\approx 0, \eeq
satisfying the constraint algebra
\begin{equation}
\Delta_{kk^{\prime}}\equiv\{\Omega_{k},\Omega_{k^{\prime}}\}
=\epsilon_{kk^{\prime}}q_{a}q_{a}\,, \label{delta}
\end{equation}
with $\epsilon_{12}=-\epsilon_{21}=1$. For the discussion to follow
it will be of central importance to convert this second-class
constraint algebra into a strongly involutive one, by suitably
embedding the model into a larger dimensional phase space.
Following the Batalin-Fradkin-Tyutin
scheme~\cite{bft,bft1,hong02pr}, we achieve this by introducing a
pair of canonically conjugate auxiliary coordinates $(\theta,
p_{\theta})$ with Poisson brackets
\begin{equation}
\{\theta, p_{\theta}\}=1. \label{phii}
\end{equation}
In this enlarged phase space one systematically constructs the
first-class constraints $\tilde{\Omega}_{i}$ as a power series in
these auxiliary coordinates, by requiring that they be in strong
involution $\{\tilde{\Omega}_{i},\tilde{\Omega}_{j}\}=0$. For the
case in question
\beq\label{Omegatilde12}
\tilde{\Omega}_{1}=\Omega_{1}+\theta,~~~
\tilde{\Omega}_{2}=\Omega_{2}-q_{a}q_{a}p_{\theta}.
\eeq
Note that
from here on all Poisson brackets are understood to be taken with
respect to the variables
$(q_{a},p_{a},\lambda,p_{\lambda},\theta,p_{\theta})$ of the
extended phase space. We next construct in the extended space the
first-class coordinates $\tilde{{\cal F}}
=(\tilde{q}_{a},\tilde{p}_{a})$, corresponding to the original
coordinates ${\cal F}=(q_{a},p_{a})$ of the second class theory.
They are again obtained as a power series in the auxiliary fields
$(\theta,p_{\theta})$ by demanding that they be in strong
involution with the first-class constraints (\ref{Omegatilde12}),
that is $\{\tilde{\Omega}_{i}, \tilde{{\cal F}}\}=0$. After some
tedious algebra, we obtain for the first-class coordinates
\begin{eqnarray}
\tilde{q}_{a}&=&q_{a}\left(\frac{q_{c}q_{c}+2\theta}{q_{c}q_{c}}\right)^{1/2}
\nonumber\\
\tilde{p}_{a}&=&\left(p_{a}+2q_{a}\lambda\frac{\theta}{q_{c}q_{c}}
+2q_{a}p_{\theta}\frac{\theta}{q_{c}q_{c}}\right)
\left(\frac{q_{c}q_{c}}{q_{c}q_{c}+2\theta}\right)^{1/2}\nonumber\\
\tilde{\lambda}&=&\lambda+p_{\theta},~~~\tilde{p}_{\lambda}=p_{\lambda}+\theta.
\label{coordinatestilde}
\end{eqnarray}
In terms of these coordinates the first-class Hamiltonian in
strong involution with the first class constraints, can be written
in the compact form
\begin{equation}
{H}_{0}(\tilde q,\tilde\lambda,\tilde
p)=\frac{1}{2}(\tilde{p}_{a}-
\tilde{q}_{a}\tilde{\lambda})(\tilde{p}_{a}-
\tilde{q}_{a}\tilde{\lambda}). \label{htilde}
\end{equation}
Notice that this is just a short hand for a Hamiltonian
$\tilde{H}_0$ which now depends on the variables
($q_a,p_a,\lambda,\theta,p_\theta$):
\beq\label{Htilde} \tilde H_0(q_a,\lambda,\theta,p_a,p_\theta)
=\frac{1}{2}\eta^2 (p_a - \lambda q_a - q_a p_\theta)^2\,, \eeq
where
\beq\label{eta}
\eta = \left(\frac{q_aq_a}{q_a q_a + 2\theta}\right)^\frac{1}{2}\,.
\eeq In terms of the first-class coordinates
(\ref{coordinatestilde}), the strongly involutive constraints
(\ref{Omegatilde12}) take the natural form \beq
\tilde{\Omega}_{1}:=\tilde{p}_{\lambda}=0,~~~
\tilde{\Omega}_{2}:=\tilde{q}_{a}\tilde{p}_{a}
-\tilde{\lambda}\tilde{q}_{a}\tilde{q}_{a}=0, \label{oott} \eeq
which thus display manifest form invariance with respect to the
second-class constraints (\ref{Omega12}). One readily checks that
one has the following Poisson brackets taken with respect to the
variables ($q_a,p_a,\theta, p_\theta,\lambda,p_\lambda)$:
\beq\label{Poissontilde1}
\{\tilde{q}_{a},\tilde{p}_{b}\}=\delta_{ab}\,,\quad
\{\tilde{q}_{a},\tilde{q}_{b}\}=0\,,\quad
\{\tilde{p}_{a},\tilde{p}_{b}\}=0,
\eeq
as well as
\bea\label{Poissontilde2}
\{\tilde{\lambda},\tilde{q}_{a}\}&=&-\frac{\tilde{q}_{a}}{\tilde
{q}_a\tilde{q}_a},~~~
\{\tilde{\lambda},\tilde{p}_{a}\}=\frac{\tilde{p}_a}{\tilde
{q}_b\tilde {q}_b} -
2\frac{\tilde{q}_a\tilde\lambda}{\tilde{q}_a\tilde{q}_a},
\nonumber\\
\{\tilde{p}_{\lambda},\tilde{p}_{a}\}&=&0\,,\quad \{\tilde\lambda,\tilde{p}_\lambda\} = 0\,,\quad
\{\tilde {p}_\lambda,\tilde{q}_a\} = 0.
\eea
From the Hamilton equations of motion we have,
using (\ref{Poissontilde1}) and (\ref{Poissontilde2}),
\bea\label{qdot-thetadot}
\dot q_a &=& \{q_a,\tilde{H}_0\} =
\eta^2(p_a - \lambda q_a - p_\theta q_a)\nonumber\\
\dot\theta &=& \{\theta,\tilde{H}_0\}
= - \eta^2(p_a - \lambda q_a - p_\theta q_a)\,.
\eea
Hence the secondary constraint $q_cq_c = const$ of the second class formulation
is replaced by $\tilde q_c\tilde q_c = q_{a}q_{a} + 2\theta = const$
in the first class formulation. This was to be expected since
the above Poisson brackets coincide with the {\it Dirac brackets} in the
original variables \cite{Girotti}. Indeed, the
``gauge invariant" variables ($\tilde q_a,\tilde
p_a,\tilde\lambda,\tilde p_\lambda$) of the first-class
formulation are just the ($q_a,p_a,\lambda,p_\lambda$) of the
second class formulation, and the corresponding Poisson brackets
correspond to the Dirac brackets, respectively. It follows from
the Poisson brackets (\ref{Poissontilde1}) and (\ref{Poissontilde2}) that the constraints
(\ref{Omegatilde12}) and the first-class Hamiltonian
(\ref{Htilde}) are all in strong involution:
\beq
\{\tilde{\Omega}_{i},\tilde{\Omega}_{j}\}=0,~~~\{\tilde{\Omega}_{i},\tilde{H}_{0}\}=0.
\label{totoij}
\eeq
This will simplify the discussion to follow.
Indeed, since $\tilde\Omega_i$ and $\tilde {H}_0$ are now in strong
involution, we can impose them strongly. Solving
$\tilde{\Omega}_{2}=0$ for $\tilde{\lambda}$, we may reduce the
Hamiltonian (\ref{htilde}) to the form \beq\label{Htilde2}
\tilde{H}_{0}=\frac{1}{2}\left(\tilde{p}_{a}
-\tilde{q}_{a}\frac{(\tilde{q}\cdot\tilde{p})}{\tilde{q}^2}\right)
\left(\tilde{p}_{a}-\tilde{q}_{a}\frac{(\tilde{q}\cdot\tilde{p})}{\tilde
{q}^2}\right). \eeq where $\tilde q\cdot\tilde p =
\tilde{q}_a\tilde{p}_a$ and $\tilde q^2 = \tilde{q}_a\tilde{q}_a
$. Effectively we have thus again only $2n$ independent degrees of
freedom in the extended phase space to describe the free particle
motion on the $(n-1)$-sphere. On the other hand, these $2n$
independent degrees of freedom satisfy the canonical Poisson algebra
(\ref{Poissontilde1}). Hence in the formulation of the Hamiltonian
(\ref{Htilde2}) we can treat the particle motion on $S^{n-1}$ as
that of an unconstrained system. This will prove very useful in
subsection 2.2.

%%%%%%%%%%%%%%%%%%%%%%%%%%%%%%%%%%%%%%%%%%%%%%%%%%%%%%%%%%%%%%%%%%%%%%%%%%
\subsection{Hamilton-Jacobi formulation}
%%%%%%%%%%%%%%%%%%%%%%%%%%%%%%%%%%%%%%%%%%%%%%%%%%%%%%%%%%%%%%%%%%%%%%%%%%
If our Lagrangian were to describe an unconstrained system, one
would have only one Hamilton-Jacobi (HJ) equation for the Hamilton
principal function $S$, as given by a partial differential
equation (PDE) of the form \beq \label{Hprime0}
H^{\pr}_{0}:=p_0+H_{0}\left(t,
q_{a},\lambda,p_a,p_\lambda\right)=0. \eeq where the substitutions
\beq \label{substitutions} {p}_{0}=\frac{\pa {S}}{\pa t},
~~~{p}_{a}=\frac{\pa S}{\pa {q}_{a}}, ~~~{p}_{\lambda}=\frac{\pa
S}{\pa {\lambda}} \eeq are understood. If $H_{0}$ does not depend
explicitly on $t$ (as in the model in question), we may then seek
a solution of the form $S=const\cdot t+W$. In the case of our
model, the primary constraint $p_{\lambda}=0$ and the secondary
constraint $\Omega_2 = q_ap_a - \lambda q_a q_a = 0$ motivate one
to consider the additional PDE \bea
H^{\pr}_1:&=& \frac{\pa S}{\pa\lambda} = 0\label{Hprime1}\\
H^{\pr}_{2}:&=&q_a\frac{\pa S}{\pa q_a}-\lambda
q_{a}q_{a}=0\label{Hprime2}. \eea It is easy to check that
equations (\ref{Hprime0}) and (\ref{Hprime1}) are inconsistent.
Thus, in the second class formulation, the above set of coupled
Hamilton-Jacobi equations admits no solution. In different terms,
they violate the integrability condition of~\cite{pi98}. Although
one can nevertheless arrive at an HJ formulation of the second
class system by making an appropriate choice of canonical
variables~\cite{Scholtz}, we circumvent the problem in the present
case by enlarging the phase space in the way described above.
%%%%%%%%%%%%%%%%
%\bigskip\noindent
%{\bf The Hamilton principal function $S$ of the extended space}
%%%%%%%%%%%%%%%%%
%\bigskip
In the extended phase space (\ref{Hprime0}), (\ref{Hprime1}) and
(\ref{Hprime2}) are replaced by
\bea
\tilde H^{\pr}_{0}:&=&\frac{\pa {\tilde S}}{\pa t } +
\frac{1}{2}\eta^2\left(\frac{\pa {\tilde S}}{\pa q_a} -
q_a\frac{\pa {\tilde S}}{\pa\theta} - \lambda q_a\right)^2.
\label{H0prime} \\
\tilde H^{\pr}_{1}:&=&\frac{\pa {\tilde S}}{\pa\lambda}+ \theta=0\label{H1prime}\\
\tilde H^{\pr}_{2}:&=&q_a\frac{\pa {\tilde S}}{\pa q_{a}} -q_a q_a
\frac{\pa {\tilde S}} {\pa\theta} - \lambda q_a q_a
=0\label{H2prime}\,.
\eea
Consider the constraint equation
$H^{\pr}_{1}=0$. It has the solution
\beq\label{f}
\tilde S(t,q_{a},\lambda,\theta) = -\frac{\alpha^2}{2} t + {\tilde
W}(q_{a},\theta) - \lambda\theta
\eeq
Hence (\ref{H0prime}) and
(\ref{H2prime}) become respectively
\bea
-\frac{\alpha^2}{2} +
\frac{1}{2}\eta^2\left(\frac{\pa {\tilde W}}{\pa q_a} -
q_a\frac{\pa {\tilde W}}{\pa\theta}\right)^2 &=& 0.
\label{H00prime} \\
q_a\frac{\pa {\tilde W}}{\pa q_{a}} -q_a q_a
\frac{\pa {\tilde W}} {\pa\theta} &=&0\label{H22prime}\,.
\eea
One possible solution of the second
equation above is
\[
{\tilde W}(q_{a},\theta) = g(q_a q_a + 2\theta) = g(\tilde q_a
\tilde q_a)
\]
with $g(x)$ so far an arbitrary function. Thus
\bea \tilde
S(t,q_{a},\lambda,\theta) &=& -\frac{\alpha^2}{2} t + g(q_{a}q_{a}
+ 2\theta) - \lambda\theta
\nonumber\\
&=& -\frac{\alpha^2}{2} t + g(\tilde q_{a}\tilde q_{a}) -
\lambda\theta
\eea
With this solution equation (\ref{H00prime})
reduces to $\alpha=0$. We now look for non-trivial solutions. Consider first
equation (\ref{H22prime}).
The relation between the variables in the first and second class formulation
motivates us to make the following Ansatz:
\beq\label{Ansatz}
\tilde W(q,\theta) = f(n\cdot\tilde q)
\eeq
with $n_{a}$ the components of a $n$-dimensional unit
vector parametrized by $n-1$ constants. Using
\beq
\frac{\partial\tilde q_c}{\partial q_a}=
\eta^{-1}\delta_{ca} - \eta \frac{2\theta}{q^2}\frac{q_a q_c}{q^2}\,,\quad
\frac{\partial\tilde q_c}{\partial\theta}=\eta\frac{q_c}{q^2}\,,
\eeq
one readily checks that equation (\ref{H22prime}) is satisfied
for any $f(x)$. This function is now determined by eq. (\ref{H00prime})
which now reads,
\[
\left(1 - \frac{(n_{a}\tilde{q}_{a})^2}{\tilde{q}_{c}\tilde{q}_{c}}\right)
f'^2(n_{a}\tilde{q}_{a}) = \alpha^2 \,.
\]
where the prime denotes the derivative with respect to the argument of $f$.
The solution to this equation has been given in \cite{Scholtz}.
Setting $x = n_{a}\tilde{q}_{a}$ and $r^2
=\tilde{q}_{a}\tilde{q}_{a}$, we thus have
\[
f'(x) = \pm\frac{\alpha}{\sqrt{1 - \frac{x^2}{r^2}}}
\]
so that upon integration in $x$ the Hamilton principal function
$\tilde W$ takes the form
\be\label{S4} \tilde W(q,\theta) =
\alpha r~\tan^{-1}\frac{n_{a}\tilde{q}_{a}}{\sqrt{r^2 -
(n_{a}\tilde{q}_{a})^2}} + const\,. \ee The Hamilton principal
function contains $n$ independent constants, which we take to be
$\alpha$ and $n_1,n_2,\ldots,n_{n-1}$, while the normalization of
$n^{a}$ implies for the $n$'th component,
$n_n=\sqrt{1-\sum_{a=1}^{n-1}n_{a}n_{a}}$. Differentiating the
Hamilton principal function with respect to these constants (new
momenta in the corresponding generating functional) yields in the
usual way \cite{Goldstein} the $n$ time-independent new
coordinates: \be\label{constantcoordinates} \beta=\frac{\partial
W}{\partial\alpha}\,,\quad \beta_a=\frac{\partial W}{\partial
n_a}\,,\quad a=1,2,\ldots n-1\,. \ee From the first equation and
(\ref{S4}) we easily obtain \be\label{ndotq}
n_{a}\tilde{q}_{a}=r\cos\frac{\beta+\alpha t}{r}\equiv
r\cos\Omega(t)\,. \ee As shown in \cite{Scholtz}, the solution
for the ``observable" $\tilde q_a$ then takes the following form in terms of $\Omega(t)$:
\beq\label{solution}
\tilde q_{a}=\frac{1}{\alpha}\left(\beta_{a}
-(n_{c}\beta_{c})n_{a}\right)\sin\Omega(t)+rn_{a}\cos\Omega(t)\,,
\eeq
where $\beta_{a}$ is the
$n$-dimensional vector
$\beta_{a}=(\beta_1,\beta_2,\ldots,\beta_{n-1},0)$. Substituting
the above result into our original condition, $r^2=\tilde{q}^2$,
leads to \be\label{r}
r=\sqrt{\frac{\beta_{a}\beta_{a}-(n_{a}\beta_{a})^2}{\alpha^2}}
\ee Thus the radius of motion is fixed if the new time independent
coordinates are specified.

%%%%%%%%%%%%%%%%%%%%%%%%%%%%%%%%%%%%%%%%%
\subsection{Schr\"odinger representation}
%%%%%%%%%%%%%%%%%%%%%%%%%%%%%%%%%%%%%%%%%
In the first-class formulation the canonical commutation relations
between the phase space variables
$(q,p),(\lambda,p_\lambda),(\theta,p_\theta)$ allows us to replace
the first class Hamiltonian (\ref{Htilde}) by the differential
operator
\beq \tilde H_0 =
\frac{1}{2}\eta(q,\theta)^2\left(\frac{\hbar}{i}\frac{\pa}{\pa
q_a}-\lambda q_a
-q_a\frac{\hbar}{i}\frac{\pa}{\pa\theta}\right)\left(\frac{\hbar}{i}\frac{\pa}{\pa
q_a}-\lambda q_a -q_a\frac{\hbar}{i}\frac{\pa}{\pa\theta}\right)
\eeq Instead of solving the corresponding eigenvalue problem in
this formulation, one may simplify this problem by noting from
(\ref{Poissontilde1}) that the variables $(\tilde q_a,\tilde p_a)$
form canonical pairs, so that we may make the replacement $\tilde
p_a \to \frac{\hbar}{i} \frac{\pa}{\pa \tilde{q}_a}$ in
(\ref{Htilde2}). In the Schr\"odinger representation, we thus have
for the quantum commutators, \beq\label{commutators}
[\tilde{q}_a,\tilde{q}_{b}] = 0\,,\quad
[\tilde{p}_a,\tilde{p}_{b}] = 0\,,\quad
[\tilde{q}_a,\tilde{p}_{b}] = i\hbar\delta_{ab}\,. \eeq Following
the symmetrization procedure of~\cite{lee81,hong99sk} we obtain
for the Hamilton quantum operator of a free particle on the
$(n-1)$-sphere (we now set $\tilde{q}^2 = r^2 = 1$)
\bea\label{Op-htilde}
\tilde{H}_{0}&=&:\frac{1}{2}\left(-i\hbar\frac{\pa}{\pa
\tilde{q}_{a}} +i\hbar\tilde{q}_{a}\tilde{q}_{c}\frac{\pa}{\pa
\tilde{q}_{c}}\right)\left(-i\hbar\frac{\pa}{\pa \tilde{q}_{a}}
+i\hbar\tilde{q}_{a}\tilde{q}_{c}\frac{\pa}{\pa
\tilde{q}_{c}}\right):
\\
&=&\frac{1}{2}\hbar^{2}\left[-\frac{\pa^{2}}{\pa \tilde{q}_{a}\pa
\tilde{q}_{a}}+(n-1)\tilde{q}_{a}\frac{\pa}{\pa
\tilde{q}_{a}}+\tilde{q}_{a}\tilde{q}_{b}\frac{\pa}{\pa
\tilde{q}_{a}}\frac{\pa}{\pa \tilde{q}_{b}}+\frac{(n+1)(n-3)}{4}
\right]\nonumber. \eea Note that the quantum Hamiltonian
(\ref{Op-htilde}) has only terms of order $\hbar^{2}$, so that one
has rotational energy contributions of order $\hbar^{2}$, without
any vibrational modes of order $\hbar$. In fact, the starting
Lagrangian (\ref{Lag}) does not possess any vibrational degrees of
freedom, since it involves only the kinetic term describing the
motions of the particle residing on the $S^{n-1}$ manifold. We now
define the Casimir operator $\tilde{J}^{2}$ in terms of the
$(n-1)$-sphere Laplacian~\cite{vil68} \beq
\tilde{J}^{2}=\hbar^{2}\left[-\frac{\pa^{2}}{\pa \tilde{q}_{a}\pa
\tilde{q}_{a}}+(n-1)\tilde{q}_{a}\frac{\pa}{\pa
\tilde{q}_{a}}+\tilde{q}_{a}\tilde{q}_{b}\frac{\pa}{\pa
\tilde{q}_{a}}\frac{\pa}{\pa \tilde{q}_{b}}\right], \label{jop}
\eeq whose eigenvalue spectrum is given in terms of the
corresponding angular quantum number $j$ $(j={\rm integers})$ and
the dimension of the sphere as follows \beq
\tilde{J}^{2}|j\rangle=\hbar^{2}j(j+n-2)|j\rangle. \label{j2op}
\eeq We thus have for the Hamiltonian operator of a free particle
on the $(n-1)$-sphere \beq
\tilde{H}_{0}=\frac{1}{2}\left[\tilde{J}^{2}+\frac{\hbar^{2}(n+1)(n-3)}{4}\right],
\label{spectrum} \eeq to yield the eigenvalue equation
$\tilde{H}_{0}|j\rangle=E_{j}|j\rangle$ with the spectrum given by
\beq
E_{j}=\frac{\hbar^{2}}{2}\left[j(j+n-2)+\frac{(n+1)(n-3)}{4}\right].\label{eigene}
\eeq Note that the Hamiltonian operator (\ref{spectrum}) is that
of a rigid rotator, and the energy eigenvalues (\ref{eigene})
involve global shifts depending on the dimension of the sphere,
which has not been included in~\cite{neves00,podo28}.
%%%%%%%%%%%%%%%%%%%%%%%%%%%%%%%%%%%%%%%%%%%%%%%%%%%%%%%%%%%%%%%%%%%%%%%%%%
\section{Gauged $O(3)$ nonlinear sigma model}
\setcounter{equation}{0}
\renewcommand{\theequation}{\arabic{section}.\arabic{equation}}
%%%%%%%%%%%%%%%%%%%%%%%%%%%%%%%%%%%%%%%%%%%%%%%%%%%%%%%%%%%%%%%%%%%%%%%%%%
In this section, we will generalize the approach developed in the
previous sections to the field theoretical $O(3)$ nonlinear sigma
model, whose Lagrangian is of the form
\begin{equation}
{\cal L}_{0}=\frac{1}{2f}(\partial_{\mu}n^{a})(\partial^{%
\mu}n^{a})+n^{0}n^{a}\pa_{0}n^{a} \label{lag}
\end{equation}
where $n^{a}$ ($a$=1,2,3) is a multiplet of three real scalar
fields which parameterize an internal space $S^{2}$, and $n^{0}$
is the Lagrange multiplier field implementing the second-class
constraint $n^{a}\pa_{0}n^{a}\approx 0$ associated with the
geometrical constraint $n^{a}n^{a}-1\approx 0$. From the
Lagrangian (\ref{lag}) the canonical momenta conjugate to the
field $n^{0}$ and the real scalar fields $n^{a}$ are given by \bea
\pi^{0}&=&0,\nonumber\\
\pi^{a}&=&\frac{1}{f}\partial_{0}{n}^{a}+n^{a}n^{0}.
\label{momenta} \eea Here one notes that $n^{0}$, $n^{a}$ and
$\partial_{0}{n}^{a}$ are entangled to define $\pi^{a}$. In terms
of the canonical momenta (\ref{momenta}), we then obtain the
canonical Hamiltonian
\begin{equation}
{\cal H}=\frac{f}{2}(\pi^{a}-n^{a}n^{0})
(\pi^{a}-n^{a}n^{0}) +\frac{1}{2f}%
(\partial_{i}n^{a})(\partial_{i}n^{a}) \label{canH}
\end{equation}
The usual Dirac algorithm is readily shown to lead to the pair of
second-class constraints $\Omega_{i}$ $(i=1,2)$ as follows \bea
\Omega_{1}&=&\pi^{0}\approx 0 \nonumber\\
\Omega_{2}&=&n^{a}\pi^{a}-n^{a}n^{a}n^{0}\approx 0.
\label{const22} \eea to yield the corresponding constraint algebra
with $\epsilon^{12}=-\epsilon^{21}=1$
\begin{equation}
\Delta_{kk^{\prime}}(x,y)=\{\Omega_{k}(x),\Omega_{k^{\prime}}(y)\}
=\epsilon^{kk^{\prime}}n^{a}n^{a}\delta^{2}(x-y). \label{delta}
\end{equation}
Following the BFT scheme~\cite{bft,bft1,hong02pr}, we
systematically convert the second-class constraints $\Omega_i=0$
$(i=1,2)$ into first-class ones by introducing two canonically
conjugate auxiliary fields $(\theta, \pi_{\theta})$ with Poisson
brackets
\begin{equation}
\{\theta(x), \pi_{\theta}(y)\}=\delta^{2}(x-y). \label{phii}
\end{equation}
The strongly involutive first-class constraints
$\tilde{\Omega}_{i}$ are then constructed as a power series of the
auxiliary fields~\cite{hong99o3},
\begin{eqnarray}
\tilde{\Omega}_{1}&=&\Omega_{1}+\theta, \nonumber \\
\tilde{\Omega}_{2}&=&\Omega_{2}-n^{a}n^{a}\pi_{\theta}.
\label{1stconst}
\end{eqnarray}
Note that the first class constraints (\ref{1stconst}) can be
rewritten as
\begin{eqnarray}
\tilde{\Omega}_{1}&=&\tilde{\pi}^{0}, \nonumber \\
\tilde{\Omega}_{2}&=&\tilde{n}^{a}\tilde{\pi}^{a}-\tilde{n}^{a}\tilde{n}^{a}
\tilde{n}^{0}, \label{oott}
\end{eqnarray}
which are form-invariant with respect to the second-class
constraints (\ref {const22}). We next construct the first-class
fields $\tilde{{\cal F}} =(\tilde{n}^{a},\tilde{\pi}^{a})$,
corresponding to the original fields defined by ${\cal
F}=(n^{a},\pi^{a})$ in the extended phase space. They are obtained
as a power series in the auxiliary fields $(\theta,\pi_{\theta})$
by demanding that they be in strong involution with the
first-class constraints (\ref{1stconst}), that is
$\{\tilde{\Omega}_{i}, \tilde{{\cal F}}\}=0$. After some tedious
algebra, we obtain for the first-class physical fields
\begin{eqnarray}
\tilde{n}^{a}&=&n^{a}\left(\frac{n^{c}n^{c}+2\theta}{n^{c}n^{c}}\right)^{1/2}\nonumber \\
\tilde{\pi}^{a}&=&\left(\pi^{a}+2n^{a}n^{0}\frac{\theta}{n^{c}n^{c}}
+2n^{a}\pi_{\theta}\frac{\theta}{n^{c}n^{c}}\right)
\left(\frac{n^{c}n^{c}}{n^{c}n^{c}+2\theta}\right)^{1/2}\nonumber\\
\tilde{n}^{0}&=&n^{0}+\pi_{\theta}\,,\quad
\tilde{\pi}^{0}=\pi^{0}+\theta, \label{pitilde}
\end{eqnarray}
and the first-class Hamiltonian (we now set right away
$\tilde{n}^2 = r^2 = 1$)
\begin{equation}
\tilde{\cal H}=\frac{f}{2}(\tilde{\pi}^{a}-
\tilde{n}^{a}\tilde{n}^{0})(\tilde{\pi}^{a}-\tilde{n}^{a}\tilde{n}^{0})
+
\frac{1}{2f}(\partial_{i}\tilde{n}^{a})(\partial_{i}\tilde{n}^{a}).
\label{htilde}
\end{equation}
Inserting the first-class constraint $\tilde{\Omega}_{2}=0$
together with $\tilde{n}^{a}\tilde{n}^{a}=1$, which are strongly
zero, into the first-class Hamiltonian (\ref{htilde}), we can
obtain $\tilde{\cal H}$ only in terms of
$(\tilde{n}^{a},\tilde{\pi}^{a})$ as follows \bea \hat{\cal H}&=&
\frac{f}{2}(\tilde{\pi}^{a}-\tilde{n}^{a}\tilde{n}^{c}\tilde{\pi}^{c})
(\tilde{\pi}^{a}-\tilde{n}^{a}\tilde{n}^{d}\tilde{\pi}^{d}) +
\frac{1}{2f}(\partial_{i}\tilde{n}^{a})(\partial_{i}\tilde{n}^{a}).
\label{htilde2} \eea Moreover, the first-class physical fields
(\ref{pitilde}) are found to satisfy the Poisson algebra
\begin{eqnarray}
\{\tilde{n}^{a}(x),\tilde{n}^{b}(y)\}&=&0, \nonumber \\
\{\tilde{n}^{a}(x),\tilde{\pi}^{b}(y)\}&=&\delta^{ab}\delta^{2}(x-y), \nonumber \\
\{\tilde{\pi}^{a}(x),\tilde{\pi}^{b}(y)\}&=&0, \label{commst}
\end{eqnarray}
which, in the extended phase space, yield the canonical quantum
commutators
\begin{eqnarray}
\left[\hat{n}^{a}(x),\hat{n}^{b}(y)\right]&=&0, \nonumber \\
\left[\hat{n}^{a}(x),\hat{\pi}^{b}(y)\right]&=&i\hbar\delta^{ab}
\delta^{2}(x-y), \nonumber \\
\left[\hat{\pi}^{a}(x),\hat{\pi}^{b}(y)\right]&=&0.
\label{commst3}
\end{eqnarray}
Note that the first-class Hamiltonian (\ref{htilde2}) does not
have extra degrees of freedom of $(\tilde{n}^{0},\tilde{\pi}^{0})$
any more so that we can have only
$(\tilde{n}^{a},\tilde{\pi}^{a})$ independent degrees of freedom
with the canonical quantum commutators (\ref{commst3}) in the
extended phase space as in the unconstrained systems.
%%%%%%%%%%%%%%%%%%%%%%%%%%%%%%%%%%%%%%%%%
\subsection{Schr\"odinger representation}
%%%%%%%%%%%%%%%%%%%%%%%%%%%%%%%%%%%%%%%%%
The quantum commutators corresponding to the Poisson brackets
(\ref{commst3}) show that we can realize the quantum operators
$\hat{\pi}^{a}$ of the O(3) nonlinear sigma model as follows: \beq
\hat{\pi}^{a}=-i\hbar\frac{\pa}{\pa\tilde{n}^{a}}. \label{op-pi}
\eeq Following the symmetrization procedure of ref.
~\cite{lee81,hong99sk}, together with ({\ref{htilde2}) and
(\ref{op-pi}), we arrive at the Hamiltonian density quantum
operator for the O(3) nonlinear sigma model \bea \hat{\cal
H}&=&:\frac{f}{2}\left(\frac{\hbar}{i}\frac{\pa}{\pa\tilde{n}^{a}}
-\frac{\hbar}{i}\tilde{n}^{a}\tilde{n}^{c}\frac{\pa}{\pa\tilde{n}^{c}}
\right) \left(\frac{\hbar}{i}\frac{\pa}{\pa\tilde{n}^{a}}
-\frac{\hbar}{i}\tilde{n}^{a}\tilde{n}^{d}\frac{\pa}{\pa\tilde{n}^{d}}
\right):
+\frac{1}{2f}(\partial_{i}\tilde{n}^{a})(\partial_{i}\tilde{n}^{a})\nonumber\\
&=&\frac{f}{2}\hbar^{2}\left(-\frac{\pa^{2}}{\pa\tilde{n}^{a}
\pa\tilde{n}^{a}}
+2\hat{n}^{a}\frac{\pa}{\pa\hat{n}^{a}}+\hat{n}^{a}\hat{n}^{b}
\frac{\pa^{2}}{\pa\hat{n}^{a}\pa\hat{n}^{b}}\right)
+\frac{1}{2f}(\partial_{i}\hat{n}^{a})(\partial_{i}\hat{n}^{a}).
\label{op-htilde} \eea Note that the Hamiltonian operator
(\ref{op-htilde}) has terms of orders $\hbar^{0}$ and $\hbar^{2}$
only, so that one has static mass (of order $\hbar^{0}$) and
rotational energy contributions (of order $\hbar^{2}$) without any
vibrational modes (of order $\hbar^{1}$). Indeed, the starting
Lagrangian (\ref{lag}) does not allow for any vibrational degrees
of freedom since it describes the motion of the soliton on the
$S^{2}$ manifold. Integrating the terms of order $\hbar^{2}$ in
the Hamiltonian operator (\ref{op-htilde}) over the
two-dimensional target manifold, one can construct the Casimir
operator $\hat{J}^{2}$ as follows \beq \hat{J}^{2}=\hbar^{2}{\cal
I}\int {\rm d}^{2}x~f\left(-\frac{\pa^{2}}{\pa\tilde{n}^{a}
\pa\tilde{n}^{a}}
+2\tilde{n}^{a}\frac{\pa}{\pa\tilde{n}^{a}}+\tilde{n}^{a}\tilde{n}^{b}
\frac{\pa^{2}}{\pa\tilde{n}^{a}\pa\tilde{n}^{b}}\right),
\label{casi} \eeq with ${\cal I}$ the moment of inertia of the
soliton (see below). Note that the above operator $\hat{J}^{2}$ is
the Laplacian on the two-sphere ~\cite{vil68} in the field
representation. Similarly, one can integrate the term of order
$\hbar^{0}$ over the two-dimensional space to define the soliton
static mass $M_{0}$ as \beq M_{0}=\int{\rm
d}^{2}x~\frac{1}{2f}(\partial_{i}\tilde{n}^{a})(\partial_{i}\tilde{n}^{a})\,.
\label{staticmass} \eeq In terms of $M_0$ and $\hat{J}^{2}$, the
Hamiltonian operator $\hat{H}$ for the O(3) nonlinear sigma model
thus takes the form: \beq \hat{H}=M_{0}+\frac{\hat{J}^{2}}{2{\cal
I}}\,. \label{hop} \eeq The associated eigenvalue problem \beq
\hat{H}|j\rangle=E_{j}|j\rangle \label{sch} \eeq leads to the
energy eigenvalues $E_{j}$ ($j=0,\pm 1,\pm 2,...$), \beq
E_{j}=M_{0}+\frac{\hbar^{2}}{2{\cal I}}j(j+1), \label{spec} \eeq
which exhibit the contribution from the static soliton mass and
the rotational excitations discussed above. ${\cal I}$ can thus be
interpreted as the moment of inertia of the soliton rigid rotator
and $j$ is the U(1) isospin quantum number associated with the
angular momentum operator $\hat{J}^{2}$ satisfying the following
eigenvalue equation in the two-dimensional space~\cite{vil68} \beq
\hat{J}^{2}|j\rangle=\hbar^{2}j(j+1)|j\rangle\label{jeigen}\,.
\eeq Note that this is the special case $n=3$ of the formula
(\ref{j2op}) in the previous section. In the O(3) nonlinear sigma
model we thus do not have a global energy shift, consistent with
the previous semiclassical result~\cite{hong99o3}. On the other
hand, in the SU(2) Skyrmion model one obtains a positive Weyl
ordering correction~\cite{hong99sk}. In the semiclassical
quantization with the ansatz \bea
\tilde{n}^{1}&=&\cos[\alpha (t)+\phi]\sin F(r) \nonumber\\
\tilde{n}^{2}&=&\sin[\alpha (t)+\phi]\sin F(r) \nonumber\\
\tilde{n}^{3}&=&\cos F(r) \label{ansatz} \eea for $\tilde{n}^{a}$
in the topological charge $Q=1$ sector, where $(r,\phi)$ are the
polar coordinates and $\alpha$ is the collective coordinate, one
can explicitly obtain the soliton mass $M_{0}$ and the moment of
inertia ${\cal I}$ as follows~\cite{hong99o3} \bea
M_{0}&=&\frac{\pi}{f}\int_{0}^{\infty}{\rm
d}rr~\left[\left(\frac{dF}{dr}\right)^{2}
+\frac{\sin^{2}F}{r^{2}}\right],\nonumber\\
{\cal I}&=&\frac{2\pi}{f}\int_{0}^{\infty}{\rm d}rr~\sin^{2}F.
\label{ei} \eea Note that the above angular momentum operator
$\hat{J}^2$ can be also constructed in the standard way from \beq
\hat{J}=\int {\rm d}^{2}x~\epsilon_{ij}x^{i}T^{0j}, \eeq where the
symmetric energy-momentum tensor is given by $T^{\mu\nu}=\frac{\pa
\tilde{\cal
L}_{0}}{\pa(\pa_{\mu}\tilde{n}^{a})}\pa^{\nu}\tilde{n}^{a}
-g^{\mu\nu}\tilde{\cal L}_{0}$, with $\tilde{\cal L}_{0}$ the
first-class Lagrangian constructed via the replacement of
$(n^{0},n^{a})\rightarrow (\tilde{n}^{0},\tilde{n}^{a})$ in the
Lagrangian (\ref{lag}). We next discuss the corresponding BRST
Lagrangean.
%%%%%%%%%%%%%%%%%%%%%%%%%%%%%%%%%%%%%%%%%
\subsection{BRST symmetries and effective Lagrangian}
%%%%%%%%%%%%%%%%%%%%%%%%%%%%%%%%%%%%%%%%%
In order to investigate the Becci-Rouet-Stora-Tyutin (BRST)
symmetries~\cite{brst} associated with the Lagrangian (\ref{lag})
of the O(3) nonlinear sigma model, we rewrite the first-class
Hamiltonian (\ref{htilde}) in terms of original fields and
auxiliary ones \beq \tilde{\cal H}=\frac{f}{2}(\pi^{a}-n^{a}n^{0}
-n^{a}\pi_{\theta})(\pi^{a}-n^{a}n^{0}
-n^{a}\pi_{\theta})\frac{n^{c}n^{c}}{n^{c}n^{c}+2\theta}
+\frac{1}{2f}(\partial_{i}n^{a})(\partial_{i}n^{a})\frac{%
n^{c}n^{c}+2\theta}{n^{c}n^{c}}, \label{hct} \eeq which is
strongly involutive with the first class constraints
$\{\tilde{\Omega}_{i},\tilde{\cal H}\}=0$. Note that with this
Hamiltonian (\ref{hct}), one cannot generate the first-class
Gauss' law constraint from the time evolution of the constraint
$\tilde{\Omega}_{1}$. By introducing an additional term
proportional to the first-class constraints $\tilde{\Omega}_{2}$
into $\tilde{\cal H}$, we obtain an equivalent first-class
Hamiltonian
\begin{equation}
\tilde{\cal H}^{\prime}=\tilde{\cal
H}+f\pi_{\theta}\tilde{\Omega}_{2}\label{hctp}
\end{equation}
to generate the Gauss' law constraint
\begin{eqnarray}
\{\tilde{\Omega}_{1}(x),\tilde{\cal
H}^{\prime}(y)\}&=&f\tilde{\Omega}_{2}\delta^{2}(x-y) , \nonumber
\\
\{\tilde{\Omega}_{2}(x),\tilde{\cal H}^{\prime}(y)\}&=&0.
\end{eqnarray}
Note that these Hamiltonians $\tilde{\cal H}$ and $\tilde{\cal
H}^{\prime}$ effectively act on physical states in the same way
since such states are annihilated by the first class constraints.
In the framework of the Batalin-Fradkin-Vilkovisky (BFV)
formalism~\cite{bfv,fik}, we now construct the nilpotent BRST
charge $Q$, the fermionic gauge fixing function $\Psi$ and the
BRST invariant minimal Hamiltonian $H_{m}$ by introducing two
canonical sets of ghost and anti-ghost fields, together with
auxiliary fields $({\cal C}^{i},\bar{{\cal P}}_{i})$, $({\cal
P}^{i}, \bar{{\cal C}}_{i})$, $(N^{i},B_{i})$ $(i=1,2)$ and the
unitary gauge choice $\chi^{1}=\Omega_{1},~~~\chi^{2}=\Omega_{2}$,
\begin{eqnarray}
Q&=&\int {\rm d}^{2}x~({\cal C}^{i}\tilde{\Omega}_{i}+{\cal
P}^{i}B_{i}),
\nonumber \\
\Psi&=&\int {\rm d}^{2}x~(\bar{{\cal C}}_{i}\chi^{i}+\bar{{\cal P}}%
_{i}N^{i}), \nonumber \\
H_{m}&=&\int {\rm d}^{2}x~\left(\tilde{\cal
H}+f\pi_{\theta}\tilde{\Omega}_{2}-f{\cal C}^{1}\bar{{\cal
P}}_{2}\right), \label{hmham}
\end{eqnarray}
with the properties $Q^{2}=\{Q,Q\}=0$ and $\{\{\Psi,Q\},Q\}=0$.
The nilpotent charge $Q$ is the generator of the following
infinitesimal transformations, \beq
\begin{array}{lll}
\delta_{Q}n^{0}=-{\cal C}^{1}, &~~\delta_{Q}n^{a}=-{\cal
C}^{2}n^{a},
&~~\delta_{Q}\theta={\cal C}^{2}n^{a}n^{a},\\
\delta_{Q}\pi^{0}=-{\cal C}^{2}n^{a}n^{a},
&~~\delta_{Q}\pi^{a}={\cal
C}^{2}(\pi^{a}-2n^{a}n^{0}-2n^{a}\pi_{\theta}),
&~~\delta_{Q}\pi_{\theta}={\cal C}^{1},\\
\delta_{Q}\bar{{\cal C}}_{i}=B_{i}, &~~\delta_{Q}{\cal C}^{i}=0, &~~\delta_{Q}B_{i}=0,\\
\delta_{Q}{\cal P}^{i}=0, &~~\delta_{Q}\bar{{\cal
P}}_{i}=\tilde{\Omega}_{i},
&~~\delta_{Q}N^{i}=-{\cal P}^{i},\\
\end{array}
\label{brstgaugetrfm} \eeq which in turn imply $\{Q,H_{m}\}=0$,
that is, $H_{m}$ in (\ref{hmham}) is the BRST invariant minimal
Hamiltonian. After some algebra, we arrive at the effective
quantum Lagrangian of the manifestly covariant form
\begin{equation}
{\cal L}_{eff}= {\cal L}_{0} + {\cal L}_{WZ} + {\cal L}_{ghost}
\label{lagfinal}
\end{equation}
where ${\cal L}_{0}$ is given by (\ref{lag}) and
\begin{eqnarray}
{\cal L}_{WZ}&=&
\frac{1}{fn^{c}n^{c}}(\partial_{\mu}n^{a})(\partial^{\mu}n^{a}){\theta}
-\frac{1}{2f(n^{c}n^{c})^{2}}\partial_{\mu}\theta\partial^{\mu}\theta,
\label{lagwz}\\
{\cal L}_{ghost}&=&-\frac{1}{2}(n^{a}n^{a})^{2}n^{0}(B+2\bar{{\cal
C}}{\cal C})
-\frac{1}{2f}(n^{a}n^{a})^{2}(B+2\bar{{\cal C}}{\cal C})^{2}\nonumber\\
&-&\frac{1}{n^{c}n^{c}}\partial_{\mu}\theta\partial^{\mu}B
+\partial_{\mu}\bar{{\cal C}}\partial^{\mu}{\cal C}.
\end{eqnarray}
This Lagrangian is invariant under the BRST transformation \beq
\begin{array}{lll}
\delta_{\epsilon}n^{0}=-2\epsilon n^{0}{\cal C},
&~~\delta_{\epsilon}n^{a}=\epsilon n^{a}{\cal C},
&~~\delta_{\epsilon}\theta=-\epsilon
n^{a}n^{a}{\cal C},\\
\delta_{\epsilon}\bar{{\cal C}}=-\epsilon B, &~~\delta_{\epsilon}{\cal C}=0, &~~\delta_{\epsilon}B=0,\\
\end{array}
\eeq where $\epsilon$ is an infinitesimal Grassmann valued
parameter.
%%%%%%%%%%%%%%%%%%%%%%%%%%%%%%%%%%%%%%%%%%%%%%%%%%%%%%%%%%%%%%%%%%%%%%%%%%
\section{Conclusion}
\setcounter{equation}{0}
\renewcommand{\theequation}{\arabic{section}.\arabic{equation}}
%%%%%%%%%%%%%%%%%%%%%%%%%%%%%%%%%%%%%%%%%%%%%%%%%%%%%%%%%%%%%%%%%%%%%%%%%%
In conclusion, it can be said that the Batalin-Fradkin-Tyutin
embedding of the second-class constrained system into a
first-class one has played a important role for obtaining the
energy spectrum, as well as a Hamilton-Jacobi formulation of the
multidimensional rigid rotator on the $S^{n-1}$ manifold. In order
to obtain its energy spectrum we made use of the Schr\"odinger
representation for this system. We have also constructed the
corresponding Hamilton principal function to obtain the nontrivial
solutions of the Euler-Lagrange equations.. We have extended these
results to a modified version of the field theoretical O(3)
nonlinear sigma model with geometric constraints, in the
Batalin-Fradkin-Tyutin scheme. Since it was possible to obtain a
Schr\"odinger realization for the quantum commutators of the
canonical fields and their conjugate momenta, we could
straightforwardly obtain the energy spectrum of the topoligical
soliton including the rotational modes, which turned out to be the
same as that of the rigid rotator on the two-dimensional target
manifold $S^2$. We have further constructed the BRST invariant
Lagrangian of this O(3) nonlinear sigma model.\\
\vskip 0.5cm STH would like to thank the warm hospitality of the
Institut f\"ur Theoretische Physik at the Universit\"at Heidelberg
during his stay, and to acknowledge financial support in part from
the Korea Science and Engineering Foundation Grant R01-2000-00015.
We also would like to thank Heinz Rothe for a helpful discussion.

\end{document}